# Investigations of Dynamic Behaviors of Lock-exchange Turbidity Currents down a Slope Based on Direct Numerical Simulation

Zhiguo He[1,2], Liang Zhao[1], Peng Hu[1,2,*], Chinghao Yu[1], and Ying-Tien Lin[1]

*Corresponding author, Email: pengphu@zju.edu.cn

[1]Ocean College, Zhejiang University, Zhoushan, 316021, China
[2]State Key Laboratory of Satellite Ocean Environment Dynamics, The Second Institute of Oceanography, State Oceanic Administration, Hangzhou 310012, China

Email: hezhiguo@zju.edu.cn (Z.H.); liangz@zju.edu.cn (L.Z.); pengphu@zju.edu.cn (P.H.); chyu@zju.edu.cn (C.Y.); kevinlin@zju.edu.cn (Y.L.)

**Abstract:** This study presents a two-dimensional (2D) direct numerical simulation (DNS) of the effects of bed slope and sediment settling on the dynamic behaviors of lock-exchange turbidity currents. The 2D DNS model is first validated against existing DNS solutions and experimental data. Afterwards, a series of numerical case studies on the effect of bed slope and sediment settling is conducted. Numerical solutions show three distinct stages of current evolution according to the behavior of the front velocity, i.e., a rapid acceleration stage due to the collapse of the dense water, followed by a second stage in which the characteristics vary with the bed slope, and finally, a deceleration stage. For a relatively steep bed (e.g., $\theta$ = 30°), the turbidity current continues to accelerate in the second stage with a reduced accelerating rate; for an intermediate bed (e.g., $\theta$ = 10°), the current appears to have a quasi-constant front velocity in the second stage; for a relatively mild slope (e.g., a flat bed), the current directly enters the final deceleration stage after the first acceleration stage. A higher settling velocity causes greater sediment settling, which reduces the driving force of the turbidity current and therefore the current front velocity. The water entrainment effect is most dominant at the very beginning of the current evolution due to the collapsing effect. Later, the water entrainment ratios drop rapidly due to the diminishing collapsing effect. Regarding the second and third stages, the bulk entrainment ratio appears to have a constant value of approximately 0.05. The absolute values of the energy components (potential energy, kinetic energy, dissipated energy, etc.) vary greatly with the bed slope and the settling velocity. The higher the bed slope is, the higher the amount of potential energy that can be converted during the turbidity current evolution. Nevertheless, when these energy components are normalized, the differences in the energy conversion patterns between the potential energy and kinetic energy are greatly reduced. Specifically, the normalized potential energy continuously decreases. The normalized kinetic energy tends to sharply increase first, then remain nearly constant, and finally decrease.
**Keywords:** Turbidity current; Direct numerical simulation; Energy; Dynamic reference plane; Front Velocity

## 1. Introduction

A turbidity current is a type of subaqueous sediment-laden gravity current [1-4], which has been studied using both a lock-exchange configuration (i.e., fixed volume) and a continuous-flux configuration. This study focuses on lock-exchange depositional turbidity currents, which represent those triggered by abrupt landslides and short-lived floods [5].

Most previous experimental results [1,6] have shown that a lock-exchange saline gravity current on flat beds may go through several well-defined phases: a relatively short acceleration phase, a slumping phase characterized by a nearly constant front velocity, and finally, a self-similar deceleration phase. Experimental results [7,8] have also indicated that the conservative gravity current first experiences an acceleration stage and then a deceleration stage. Direct Numerical Simulations (DNS) [9,10] show a nearly constant stage between the acceleration and deceleration stages. The situation is even more complex in regard to the non-conservative turbidity current down a slope [11]. Essentially, the current front velocity is controlled by a driving force (i.e., the negative buoyancy force) that arises from the density contrast between the current and the ambient water. The density contrast is affected by the water entrainment at the upper interface and sediment exchange at the topography. An understanding of sediment deposition has been well-established (see Eq. (3) in Section 2). In contrast, quantification of water entrainment receives relatively little study. Previous empirical relations on water entrainment ratios [12,13] are mostly derived using experimental data from conservative gravity currents, in particular within the continuous-flux configuration. For those experimental studies within the lock-exchange configuration, most results [7,8,14] are presented in the form of the bulk values, i.e., the duration-averaged entrainment ratio. Hallworth *et al.* [15] determined an average entrainment ratio of the lock-exchange gravity current on a flat bed as 0.063 ± 0.003. ,

in which the entrainment ratio is defined as the ratio of the volumes of ambient and original fluid in the head. Recently, both the experimental and large eddy simulation results [16-18] indicate that the entrainment ratio of a lock-exchange gravity current is closely related to the initial density difference and the aspect ratio of the initial water depth to the lock length. However, the values of the water entrainment of a gravity current developing on a slope are controversial. For example, Krug *et al.* [13] experimentally determined the entrainment ratio of a constant-influx gravity current up a 10° slope as around 0.04, whereas the computational values from large eddy simulations [19] show that the entrainment ratio decreases with a larger negative-slope angle and the value is around 0.02 at a 3.5° negative-slope. For a turbidity current down a positive-slope, the effects of sediment and slope on water entrainment are still unclear. The temporal and spatial variations of the water entrainment, and the effects of suspended particles and bottom topography are still needed to be evaluated.

To shed insight on the abovementioned issues, high-resolution numerical simulations are one of the best choices. Various numerical models are feasible, including Reynolds-averaged Navier-Stokes simulation (RANS) [20-23], Large-Eddy Simulation (LES) [11,24-27], and DNS [28-31] (A detailed review can be found in Meiburg *et al*. [1].). The DNS approach is applied here because it can provide the most accurate information on the flow field. Adoption of the DNS solver is further motivated by its advantages to analyze the energy budget [29]. The 2D DNS has made great contributions to elucidate the physics of fluids, such as the self-sustaining condition of turbidity currents [32], instabilities in a near-critical fluid layer [33], and the effect of small-scale forcing on large-scale structures [34]. Blanchette *et al.* [32] have applied 2D DNS to investigate the conditions for self-sustainment of turbidity currents. Prof. Ungarish and his colleagues [35, 36, 37] have used 2D DNS to validate theoretical models, such as in the exchange of energy [35,36] and the front velocity [37-39]. This study aims to reveal the effects of the particle settling velocity and bed slope on the bulk properties of particle-laden gravity currents (e.g., the longitudinal variation in front velocity, energy budget and entrainment effect); instead of resolving the three-dimensional (3D) structures, the 2D DNS is used.

This paper is organized as follows. First, the numerical model, including the governing equations, boundary conditions, and numerical scheme, is introduced. Then, the model is validated by comparing the simulation results with data in the published literature. Later, we present detailed analyses of the dynamics behaviors of the current obtained from the DNS model. Finally, conclusions are drawn.

# 2. Two-dimensional DNS solver for depositional turbidity currents

## 2.1. Problem description

FIG. 1 shows the sketch for the lock-exchange turbidity current configuration. The turbidity current develops in a rectangular flume (height: $\hat{H}$; length: $\hat{L}$), with a lock gate separating the initial sediment-laden water and clear water. A reservoir storing turbid water (length: $\hat{l}$; height: $\hat{h}$; sediment concentration: $\hat{c}_0$) is located on the left side of the lock gate. On the right side of the lock gate is a slope, with $\theta$ denoting the slope angle, above which clear water is stored. The water level of the clear-water part of the flume is the same as that of the turbid water. The horizontal and vertical directions are denoted as $\hat{x}$ and $\hat{z}$, respectively. Upon the lock-releasing, the turbid water would intrude to the right and propagate along the slope. While moving downstream, sediment deposits on the flume bed, reducing the density and thus the driving force of the current. Gradually, the current slows down and will finally die out. In FIG. 1, the distance $\hat{X}_f$ measures the distance between the current front to the slope transition position. The current front position is determined by a threshold value of the particle concentration. Here, a nondimensional threshold concentration of 0.01 is used, and sensitivity analysis indicates that the front location is nearly invariable when the threshold value varies between 0.01 and 0.1. The sign ˆ refers to a dimensional quantity.

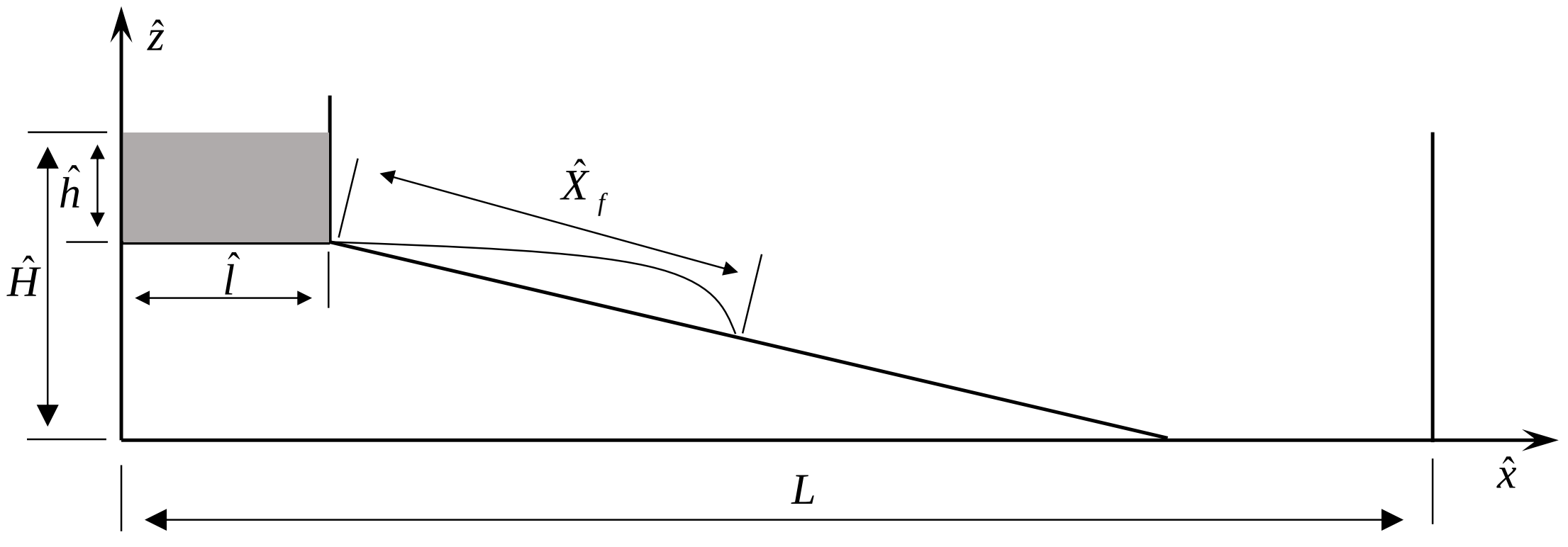

FIG. 1. Sketch view of the setup for a lock-exchange turbidity current down a slope.

## 2.2. Two-dimensional Navier-Stokes governing equations

An analysis of DNS numerical solutions commonly adopts dimensionless parameters. Using the half height $h/2$ as the length scale and the initial buoyancy velocity $\hat{u}_b$ as the velocity scale ($\hat{u}_b = \sqrt{\hat{h}\hat{g}\hat{c}_0(\hat{\rho}_p - \hat{\rho})/\hat{\rho}/2}$), the *Re* number can be defined as $Re = \hat{u}_b\hat{h}/(2\hat{\nu})$. Here, $\hat{\nu}$ is the kinematic viscosity of the fluid; $\hat{\rho}_p$ and $\hat{\rho}$ are the density of the sediment and the fluid, respectively; $g$ is gravitational acceleration; and $\hat{c}_0$ is the initial volume fraction of the sediment. Other scales include the following: $\hat{h}/(2\hat{u}_b)$ is the time scale; $\hat{c}_0$ is the sediment concentration scale; and $\hat{\rho}\hat{u}_b^2$ represents the pressure scale. With these scales, nondimensional parameters can be defined and are summarized in Table 1.

Table 1. The dimensional and nondimensional parameters.

| Name | Dimensional form | Nondimensional form |
|---|---|---|
| Length of domain | $\hat{L}$ | $L = \frac{\hat{L}}{\hat{h}/2}$ |
| Height of domain | $\hat{H}$ | $H = \frac{\hat{H}}{\hat{h}/2}$ |
| Length of initial dense fluid | $\hat{l}$ | $l = \frac{\hat{l}}{\hat{h}/2}$ |
| Height of initial dense fluid | $\hat{h}$ | $h = \frac{\hat{h}}{\hat{h}/2} = 2$ |
| Front location | $\hat{X}_f$ | $X_f = \frac{\hat{X}_f}{\hat{h}/2}$ |
| Horizontal length | $\hat{x}$ | $x = \frac{\hat{x}}{\hat{h}/2}$ |
| Vertical length | $\hat{z}$ | $z = \frac{\hat{z}}{\hat{h}/2}$ |
| Horizontal velocity of fluid | $\hat{u}_x$ | $u_x = \frac{\hat{u}_x}{\hat{u}_b}$ |
| Vertical velocity of fluid | $\hat{u}_z$ | $u_z = \frac{\hat{u}_z}{\hat{u}_b}$ |
| Front velocity | $\hat{U}_f$ | $U_f = \frac{\hat{U}_f}{\hat{u}_b}$ |
| Time | $\hat{t}$ | $t = \dfrac{\hat{t}}{\hat{h}/(2\hat{u}_b)}$ |
| Pressure | $\hat{p}$ | $p = \dfrac{\hat{p}}{\hat{\rho}\hat{u}_b^2}$ |
| Settling velocity | $\hat{u}_s$ | $u_s = \dfrac{\hat{u}_s}{\hat{u}_b}$ |
| Sediment concentration | $\hat{c}$ | $\hat{c} = \dfrac{\hat{c}}{\hat{c}_0}$ |

Using these nondimensional parameters and considering the situation of relatively low sediment concentration, the 2D governing equations are [40]

$$\frac{\partial u_x}{\partial x} + \frac{\partial u_z}{\partial z} = 0, \tag{1}$$

$$\frac{\partial u_x}{\partial t} + u_x \frac{\partial u_x}{\partial x} + u_z \frac{\partial u_x}{\partial z} = -\frac{\partial p}{\partial x} + \frac{1}{Re}\left(\frac{\partial^2 u_x}{\partial x^2} + \frac{\partial^2 u_x}{\partial z^2}\right), \tag{2a}$$

$$\frac{\partial u_z}{\partial t} + u_x \frac{\partial u_z}{\partial x} + u_z \frac{\partial u_z}{\partial z} = -\frac{\partial p}{\partial z} + \frac{1}{Re}\left(\frac{\partial^2 u_z}{\partial x^2} + \frac{\partial^2 u_z}{\partial z^2}\right) - c, \tag{2b}$$

$$\frac{\partial c}{\partial t} + u_x \frac{\partial c}{\partial x} + (u_z - u_s)\frac{\partial c}{\partial z} = \frac{1}{ReSc}\left(\frac{\partial^2 c}{\partial x^2} + \frac{\partial^2 c}{\partial z^2}\right) \tag{3}$$

where $u_x$ and $u_z$ are the fluid velocity in the $x$ and $z$ direction, respectively; $t$ is the time; $p$ is the pressure; $c$ is the nondimensional volume fraction of sediment; $u_s = \hat{d}_p^2(\hat{\rho}_p - \hat{\rho})\hat{g}/(18\hat{\rho}\hat{\nu})/\hat{u}_b$ is the nondimensionalized setting velocity of the sediment; $Sc = \hat{\nu}/\hat{k}$ is the Schmidt number ($\hat{k}$: the diffusivity of the sediment concentration field), which is set equal to unity here [40]; and $\hat{d}_p$ is the diameter of sediment.

### 2.3 Boundary conditions

A free-slip condition is applied at the top boundary (i.e., $z = H$). A no-slip condition is applied at the bottom topography and sidewalls (i.e., $x = 0, L$), where $H$ and $L$ are the nondimensional height and length of the computational domain, respectively. A no-flux boundary condition is implemented at the sidewalls and the top boundary [40]:

$$\frac{\partial c}{\partial x} = 0, \quad x = 0, L. \tag{4}$$

$$u_s c + \frac{1}{ScRe}\frac{\partial c}{\partial z} = 0\,, \quad z = H\,. \tag{5}$$

At the bottom topography, the particle is assumed to leave the flow due to sedimentation, which is numerically achieved by [29]

$$\frac{\partial c}{\partial t} = u_s \cos\theta \frac{\partial c}{\partial z_{\perp}}\,. \tag{6}$$

where $z_{\perp}$ denotes the direction of the normal vector on the bottom surface.

## 2.4 Numerical method

The initial nondimensional concentration field is prescribed as unity in the gray area (FIG. 1) and zero elsewhere. A distance function is applied at the interface, which is smoothed by solving the initialization equation [41]. The convection and diffusion terms in the transport equation are approximated using the upwinding combined compact difference (UCCD) algorithm [42,43]. The time integration for the transport equation is performed via a third-order Runge-Kutta (TVD-RK3) scheme [44] to calculate the particle concentration $c'$. The convection terms in the momentum equation are discretized using the third-order quadratic upwind interpolation for convective kinematics (QUICK) scheme, and the diffusion terms are approximated by the second order center difference scheme. Finally, to obtain the velocity field, a projection method [45] is applied: an intermediate velocity $u^*$ is first computed by ignoring the pressure gradient term and the virtual force term. The pressure Poisson equation is then solved using the Gauss-Seidel iteration solver to obtain the pressure field, which is then used to correct the intermediate velocity (i.e., compute $u'$). Then, the immersed boundary method [46] is applied to consider the effects of the topography such that $u'$ and $c'$ are further corrected (i.e., compute $u^{n+1}$ and $c^{n+1}$) depending on the different locations of the mesh cell. The immersed boundary method is implemented by adding a virtual force in the momentum equation. The solid phase acts on the fluid phase by the virtual force expressed by

$$f^{\,n+1} = \frac{u^{n+1} - u'}{\eta \Delta t}\,, \tag{7}$$

where $\Delta t$ is the time step and $\eta$ is the percent of the volume of the solid phase (slope and the area below) in one mesh cell and is calculated by further dividing the cell into finer grids, as shown in FIG. 2. If $\eta = 0$, the mesh cell is completely in the region that is outside the complex topography, i.e., the slope; thus, $c^{n+1} = c'$ (at $(x_{k,q}, z_{k,q})$), and $u^{n+1} = u'$ (at $(x_{k,q}, z_{k,q})$). If $\eta \neq 0$, the boundary of the topography crosses the mesh cell, or the mesh cell is completely in the region of the complex topography; thus, $c^{n+1} = c'(1-\eta)$ (at $(x_{k,q}, z_{k,q})$), and $u^{n+1} = u'(1-\eta)$ (at $(x_{k,q}, z_{k,q})$).

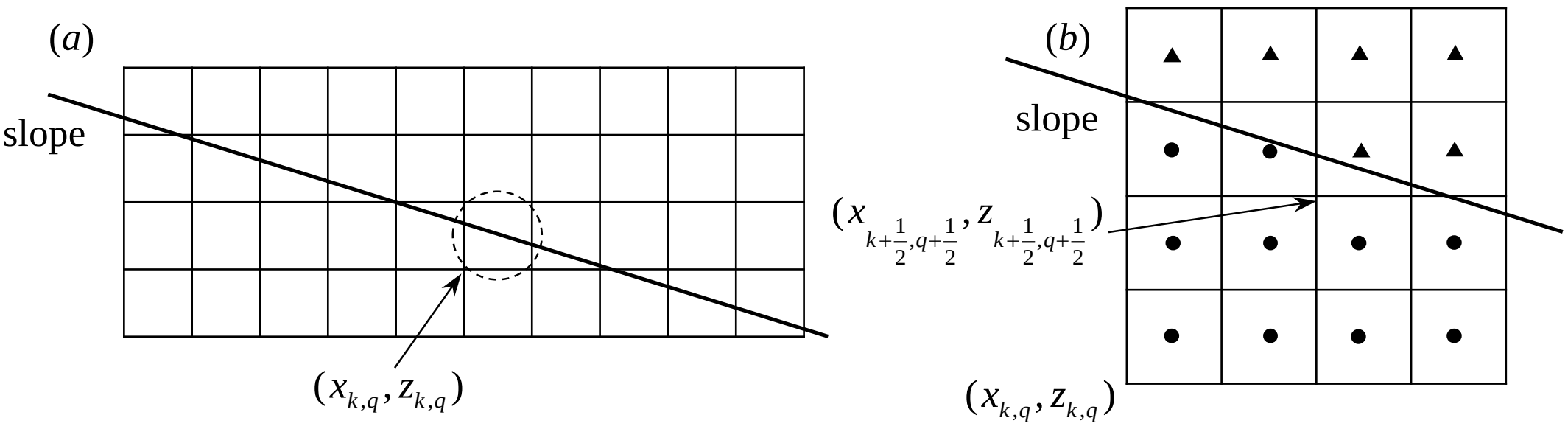

FIG. 2. ($a$) Uniform mesh cells in Cartesian grids. The dotted circle is magnified in ($b$). The mesh cell is further divided into finer grids. We only show 16 finer grids here. ● and ▲ are located in the center of the finer grid. ●: The finer grid belonging to the solid phase. ▲: The finer grid belonging to the fluid phase. In this case, $\eta$ = 10/16. Judgment of the mesh cell location is performed before the calculation.

## 2.5 Validation of the numerical model

Here, the DNS numerical solver is validated by comparing the computed front location against both the previous numerical solutions and the experimental data. We first present a comparison of the front location with the simulation results of Necker *et al.* [29], as shown in FIG. 3. The simulation case is conducted on a flat bed, with dimensionless parameters of $L$ = 18, $h$ = 2, $l$ = 1, $u_s$ = 0.02, and $Re$ = 2236. 1800 × 200 mesh cells are used. A good agreement can be seen in FIG. 3.

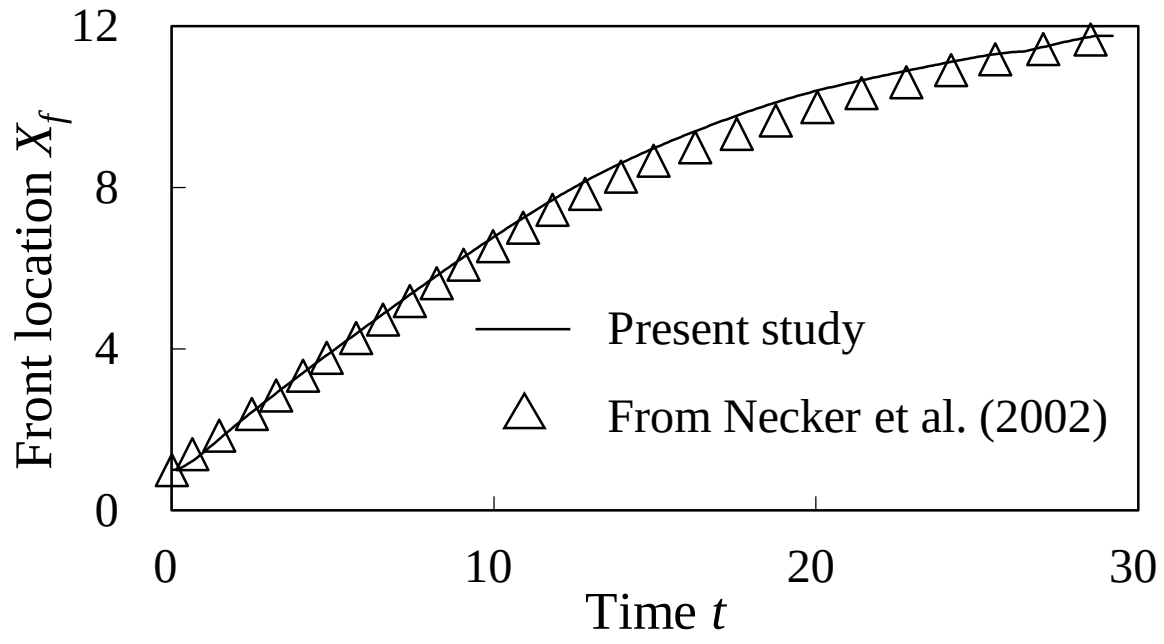

FIG. 3. Comparison of $X_f$ from the simulation case of Necker *et al.* [29] with that in the present study.

We proceed with a comparison with experimental data of B1 reported by Kubo

[47], of which the main parameters are tan($\theta$) = 0.1, $\hat{L}$ = 10 m, $\hat{l}$ = 0.5 m, $\hat{h}$ = 0.2 m, $\hat{H}$ = 0.4 m, $\hat{c}_0$ = 0.02, $\hat{u}_s$ = 0.0055 m/s, and $\hat{\rho}_p$ = 2650 kg/m$^3$. The corresponding dimensionless parameters are $L$ = 100, $h$ = 2, $l$ = 5, $H$ = 4, $u_s$ = 0.03, and $Re$ = 17865. The computational length $L$ is set to 26 in the simulations. A total of 3510 × 540 cells are used. FIG. 4 also includes the measured data, which is reproduced satisfactorily by the numerical solver.

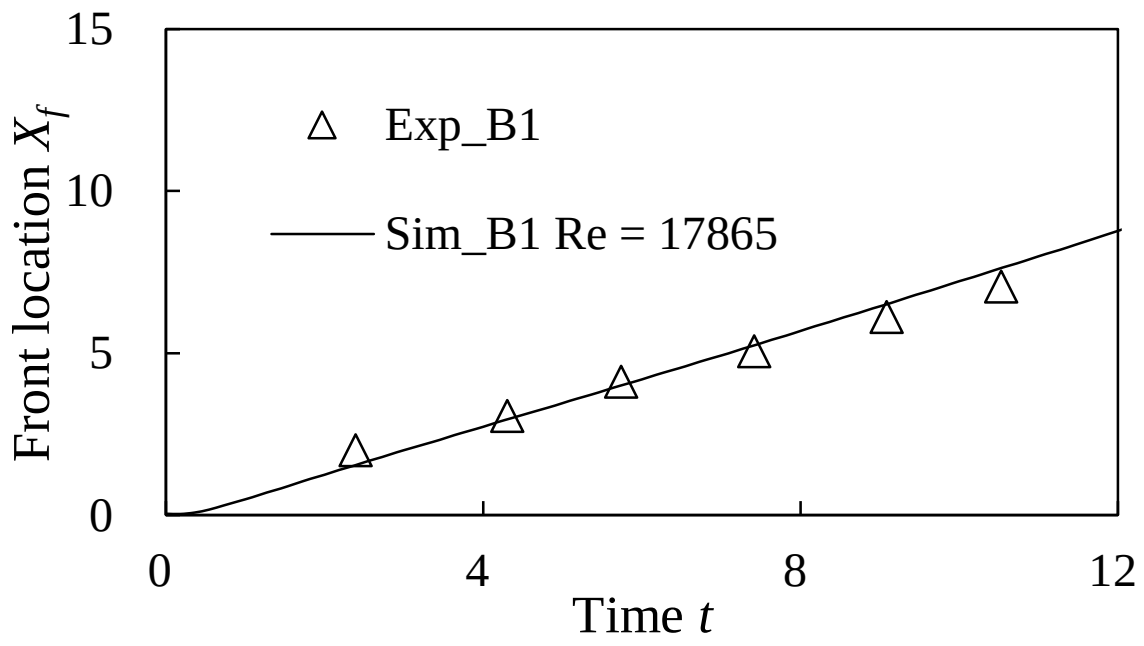

FIG. 4. Comparisons of the front location measured experimentally by Kubo [47] with the data in our simulation within 12 nondimensional time units.

## 3 Numerical case studies: dynamic behaviors of turbidity currents

As summarized in Table 2, seven numerical cases are designed with slopes ranging from 0° to 30° and sediment settling velocity ranging from 0 to 0.02. The computational domain (i.e., $L$ × $H$) is set sufficiently large such that within the computational time, the turbidity current does not arrive at the boundary of the domain. Here, $h/l$ is set equal to 2. The Reynolds number $Re$ is set to 3000. The horizontal and vertical spatial step $\Delta x$ is set to 2/110, which is of a similar order as $O(1/(ReSc)^{0.5})$ and thus satisfies the requirement for high-quality DNS modeling [9]. The time step $\Delta t$ is set to 0.005$\Delta x$.

One of the advantages of applying dimensionless parameters is that one dimensionless simulation can represent many dimensional cases. For example, in a simulation case, the dimensionless height of the initial dense fluid is $h$ = 2. If the characteristic length scale is 1 cm, the dimensionless simulation represents a case with $\hat{h}$ = 2 cm. However, if the characteristic length scale is 1 m, the dimensionless simulation represents a case with $\hat{h}$ = 2 m. The dimensional scale varies with the characteristic scale. Here, we take the nondimensional parameters of case 1 in Table 2 as an example to express the represented dimensional scales. The nondimensional parameters of case 1 are $L$ = 16, $H$ = $h$ = 2, $l$ = 1, $Re$ = 3000, $u_s$ = 0.02 and $\theta$ = 0. The corresponding dimensional scales can be $\hat{L} = 1.28\ m$, $\hat{H} = \hat{h} = 0.16\ m$, $\hat{l} = 0.08\ m$, $\hat{c}_0 = 0.1\%$, $\hat{d}_p = 30\ \mu m$, $\hat{\rho}_p = 2650\ kg/m^3$, $\hat{\rho} = 1000\ kg/m^3$, $\hat{u}_s = 0.008\ m/s$, and $\theta$ = 0.

Table 2. Simulation cases in this study

| Case | 1 | 2 | 3 | 4 | 5 | 6 | 7 |
|---|---|---|---|---|---|---|---|
| Slope (°) | 0 | 6 | 10 | 10 | 10 | 15 | 30 |
| $u_s$ | 0.02 | 0.02 | 0.02 | 0 | 0.005 | 0.02 | 0.02 |

### 3.1 Development of the turbidity current

The computed current evolution process from case 3 is shown in FIG. 5. Once the lock is released, the heavy sediment-laden fluid collapses down immediately. Within the initial short period, the evolution of the turbidity current is mainly controlled by this collapsing effect. Later, as the dense fluid continues to flow down, a clear head structure is developed from $t \approx 1.4$. At the lower interface between the turbidity current and the slope, a lifted nose is formed. At the upper interface between the turbidity current and the ambient water, a velocity-shear layer is generated, which promotes the Kelvin–Helmholtz (KH) instabilities and turbulent billows. In addition, as time passes, a raising cloud emerges at the tail part of the current. FIG. 5 (*g*) and (*h*) show the velocity field and the streamlines of the turbidity current at $t$ = 5.72, respectively. It can be identified that the velocity vectors and the streamlines appear uniformly in the main body of the current. However, in the area with the KH instabilities, the turbulent billows, and the raising cloud, the velocity vectors grow irregularly, and smaller streamline circles develop. These structures generate mixing at the upper interface, and the ambient clear water is entrained into the descending current. The density of the turbidity current is persistently reduced by water entrainment as well as settling of suspended sediments. These observations are consistent with [9,32], suggesting that the present model has satisfactorily reproduced development of a turbidity current.

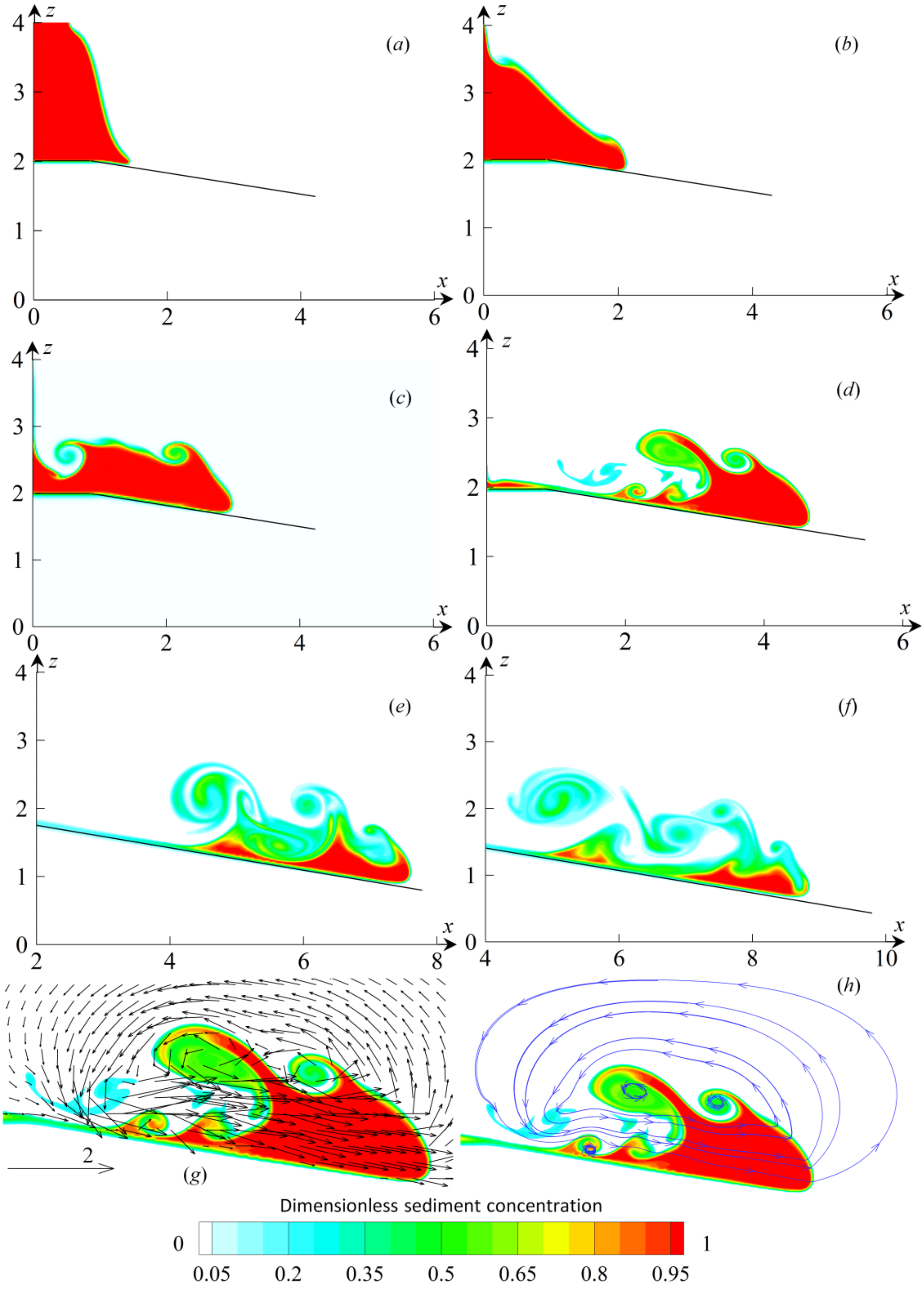

FIG. 5. Time evolution of a turbidity current down a 10° slope. *Re* = 3000. The turbidity current is visualized by the contour of the sediment concentration. (*a*) $t = 1$; (*b*) $t = 1.4$; (*c*) $t = 3.27$; (*d*) $t = 5.72$; (e) $t = 10$; (*f*) $t = 12$. The velocity field (vector) and streamline (blue line) of the turbidity current at $t = 5.72$ are shown in (*g*) and (*h*), respectively.

FIG. 6 shows the comparison of turbidity currents down different slopes at $t = 6$.

The typical characteristics of the abovementioned turbidity current still exist as the slope varies. Although a recent study [48] indicated that the KH instabilities and turbulent billows could be greatly damped when $\theta$ = 0°, these turbulent structures have clearly been seen in many previous experimental and numerical studies [1,28,29,49] of compositional gravity currents and turbidity currents on a flat bed. As the slope angle increases, the length of the turbidity current head becomes shorter and its height becomes larger. In addition, the two main turbulent billows at the upper interface tend to become more complex. When the slope angle increases to 30°, the two billows are even mixed together, and a larger cloud is then formed. The change in the shape indicates that the mixed region of the turbidity current increases as the slope angle increases. To demonstrate this conclusion quantitatively, the mixed region of the turbidity current $A_m$ is calculated, which is defined as the area where the particle concentration is in the range between 0.01 and 0.99 in the whole domain, i.e.,

$$A_m = \int_0^H \int_0^L \phi_{(c)} dxdz, \quad \text{with} \begin{cases} \phi_{(c)} = 1, & 0.01 < c < 0.99 \\ \phi_{(c)} = 0, & c \le 0.01 \text{ or } c \ge 0.99 \end{cases}, \tag{8}$$

where $\phi_{(c)}$ is a defined parameter changing with $c$. The time evolution of the normalized volume of the mixed region in different cases is presented in FIG. 7. At the initial period during which the turbidity current is mainly controlled by the collapsing process, the mixed region is not sensitive to the change in slope angle. While at the later stage, the increase in the slope angle makes the area of the mixed region grow more quickly. This conclusion agrees with the previous numerical study of the compositional gravity current [9]. However, it is more appropriate to evaluate the mixing and entrainment effect between the current and the ambient water by the dimensionless parameter, i.e., the entrainment ratio [50], which is related to both the entrainment volume of the ambient water and the front velocity of the current. The analysis of the front velocity can also serve as a quantitative approach to show the development process of the turbidity current. In Section 3.3, a detailed analysis of the front velocity is presented.

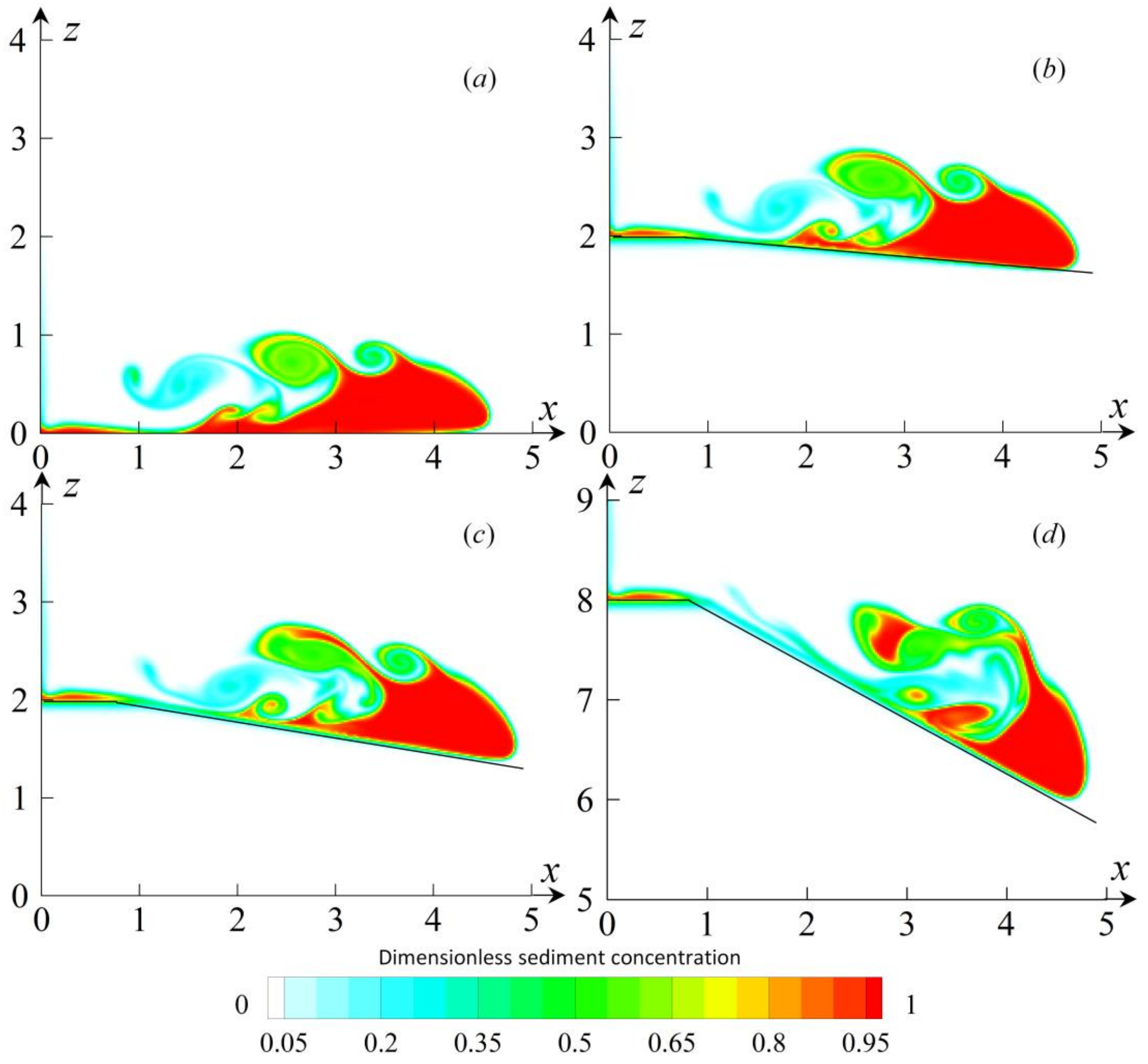

FIG. 6. Turbidity currents down different slope angles at $t$ = 6. ($a$) $\theta$ = 0°; ($b$) $\theta$ = 6°; ($c$) $\theta$ = 10°; ($d$) $\theta$ = 30°. The turbidity current is visualized by the contour of the particle concentration.

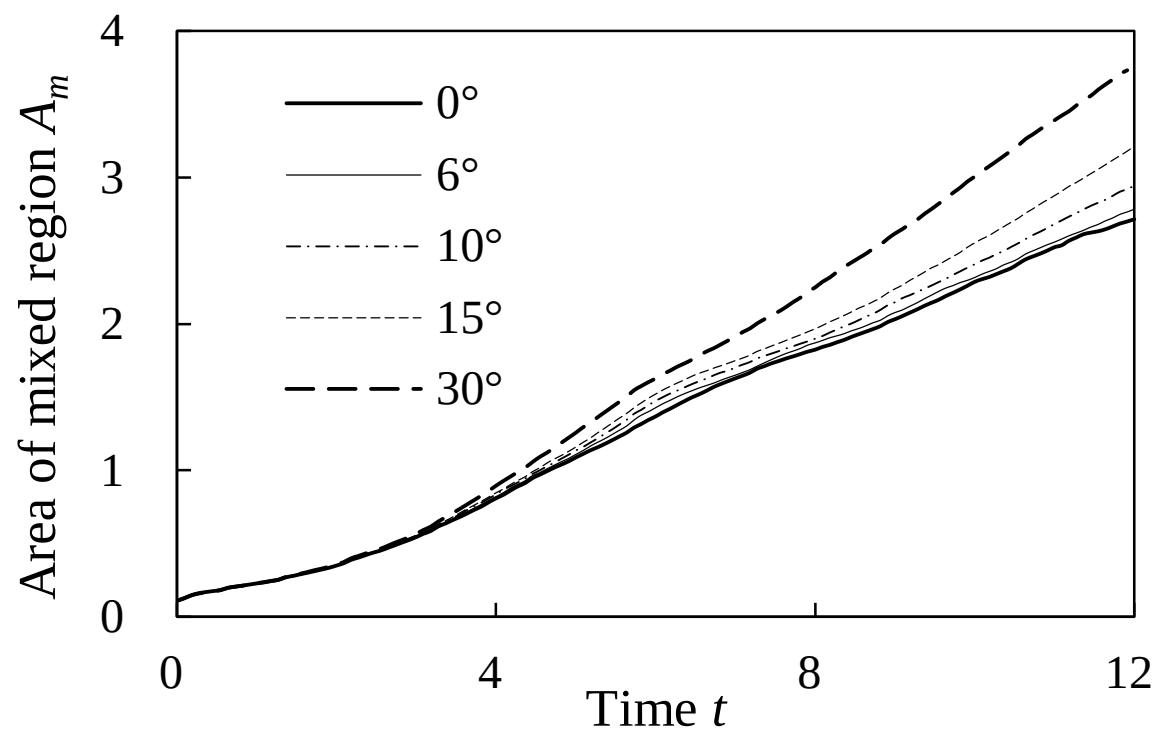

FIG. 7. Time evolutions of the mixed region of turbidity currents down different slope angles. $Re$ = 3000. The values are normalized by the initial area of the turbidity current.

### 3.2 Energy budget

Turbidity current evolution is essentially an energy conversion process. The initial potential energy of the turbid water behind the lock, i.e., $E_{p0}(t)$, is the energy source. During evolution, the initial potential energy will be transformed into three parts: 1) those that have been converted to kinetic energy, i.e., $E_{con}(t)$; 2), those that become temporally unconvertible due to sediment deposition (this part may become convertible if resuspension is allowed; however, this is not the present focus), i.e., the inactive potential energy $E_d(t)$; and 3) those remaining as active potential energy $E_p(t)$. Specifically, 'active/inactive' describe whether this part of the potential energy can be further transformed into kinetic energy or not. The converted kinetic energy $E_{con}(t)$ has two elements: 1) those remaining as the kinetic energy during the current evolution, i.e., $E_k(t)$, and 2) those $E_{dis}(t)$ that have been dissipated. Therefore [40]

$$\underset{\text{Total potential energy}}{\underset{\downarrow}{E_{p_0}(t)}} = \overset{\text{PE that is not converted}}{\overset{\uparrow}{E_r(t)}} + \underset{\text{PE that has not been converted into KE}}{\underset{\downarrow}{E_{con}(t)}} = \underbrace{\overset{\text{PE that can be further converted}}{\overset{\uparrow}{E_p(t)}} + \overset{\text{PE that becomes inactive due to deposition}}{\overset{\uparrow}{E_d(t)}}}_{E_r(t)} + \overbrace{\underset{\text{KE excluding dissipated part}}{\underset{\downarrow}{E_k(t)}} + \underset{\text{Dissipated KE}}{\underset{\downarrow}{E_{dis}(t)}}}^{E_{con}(t)} . \quad (9)$$

For description convenience, the remaining potential energy including both convertible and unconvertible parts is defined as $E_r(t) = E_p(t) + E_d(t)$.

Necker *et al.* [29,30] presented an analysis of the energy budget for lock-exchange turbidity currents on a flat bed. Ouillon *et al.* [51] extended it to turbidity currents over slopes. The specific expressions for these energy components have been derived for the flat bed by Necker *et al.* [29,30] and for sloping beds by Ouillon *et al.* [51]. They are reformulated as follows

$$E_k(t) = \int_0^H \int_0^L \frac{1}{2} uudxdz, \quad (10a)$$

$$E_p(t) = \int_0^H \int_0^L c(z - z_d)dxdz, \quad (10b)$$

$$E_d(t) = \int_0^L m_d (z_b - z_d)dx, \quad (10c)$$

$$E_{dis}(t)=\int_0^t\left(\varepsilon_f(\tau)+\varepsilon_s(\tau)\right)d\tau\,, \tag{10d}$$

where $\varepsilon_f(t)=\int_0^H\int_0^L \frac{2}{Re} ssdxdz$, and $\varepsilon_s(t)=\int_0^H\int_0^L u_s cdxdz$ are the energy dissipation rates; $s=\frac{1}{2}(\frac{\partial u_x}{\partial z}+\frac{\partial u_z}{\partial x})+\frac{1}{2}(\frac{\partial u_z}{\partial x}+\frac{\partial u_x}{\partial z})+\frac{1}{2}(\frac{\partial u_x}{\partial x}+\frac{\partial u_x}{\partial x})+\frac{1}{2}(\frac{\partial u_z}{\partial z}+\frac{\partial u_z}{\partial z})$ is the rate-of-strain tensor; $z_d$ is a reference plane; $z_b$ is the bed elevation; and $m_d=\int_0^t u_s c(x,z_b,\tau)d\tau$ is the volume of deposited sediments per unit area;

From Eqs. (10b) and (10c), the potential energy components $E_p(t)$ and $E_d(t)$ depend on the reference plane, i.e., $z_d$. Necker *et al.* [29,30] defined the reference plane at the flat bed. This makes $E_d(t)$ invariably equal to zero and explains the lack of the term '$E_d(t)$' in Necker *et al.* [29,30]. For a turbidity current down a slope, the term $E_d(t)\neq 0$ because sediments deposit at different heights. From a physical perspective, the idealized reference plane $z_d$ must be defined at the position where the current finally dies out. However, it is difficult to define the time that the current dies out because the mixing process lasts for a very long time [31]. As a compromise, this study proposes a dynamic reference plane: at a certain time *t*, the reference plane is defined at the lowest point of the turbidity current, which is determined by a threshold value of the sediment concentration of 0.01. In the following, analyses of the energy terms are conducted against time by following the conventional practice [9,29].

### 3.2.1 Influence of the slope angle on the energy budget

FIG. 8 presents the temporal variations of the kinetic energy $E_k$ for turbidity currents down different slopes (cases 1, 2, 3, 6, and 7). At the initial period ($t < 1.6$), the kinetic energy $E_k$ experiences a rapid increase. This is because at the initial stage, the flow is mainly controlled by the collapsing effect. As the current moves forward, the collapsing effect diminishes, and the slope effects accumulate. Consequently, considerable differences are seen after the initial stage. On a flat bed (case 1), the kinetic energy $E_k$ starts to decrease. On sloping beds (cases 2, 3, 6 and 7), the kinetic energy $E_k$ continues to increase, although at a reduced rate. A gentler slope tends to exhibit a slower increase in the kinetic energy, which is also due to the slope effects. The steeper the bed slope, the larger the amount of potential energy that can be converted. Nevertheless, as sediment continues to deposit, it is expected that the kinetic energy over the largest slope (e.g., $\theta$ = 30°) would also decrease after propagating a sufficiently long distance, though this phase is not simulated in this paper.

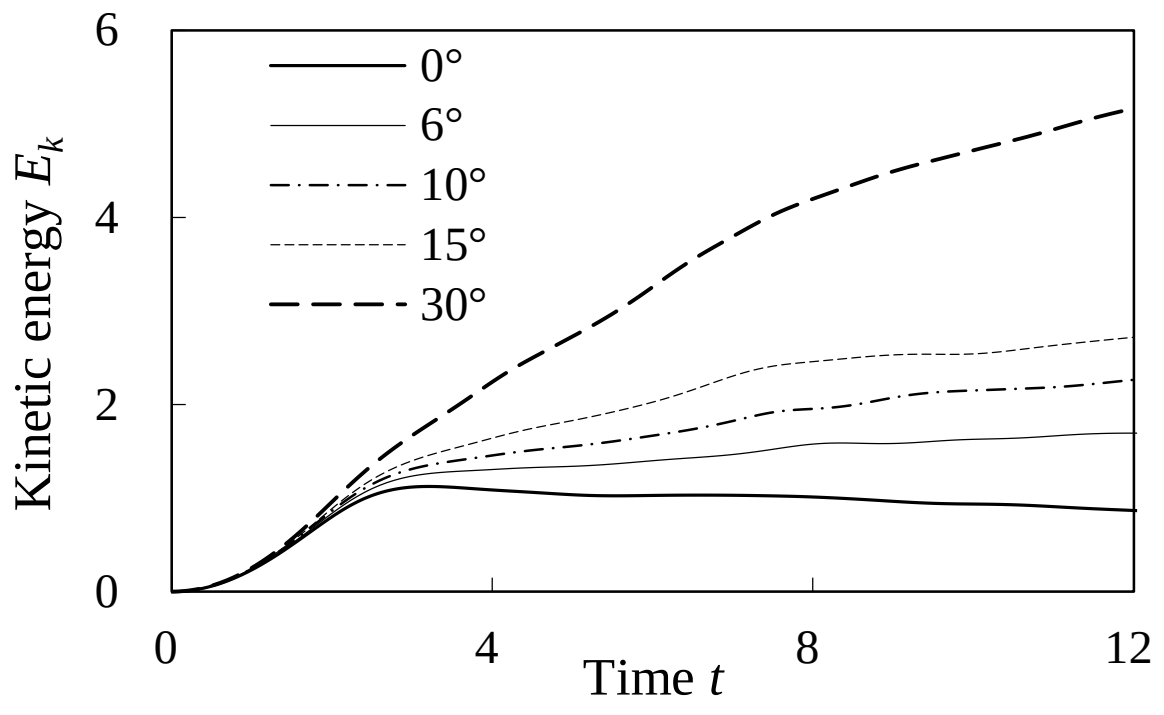

FIG. 8. Time evolutions of $E_k$ for turbidity currents down different slopes.

We proceed to discuss the proportions of different energy components within the total energy. FIG. 9 shows the time evolutions of the proportions of the remaining kinetic energy $E_k$ (FIG. 9 (*a*)), the active potential energy $E_p$ (FIG. 9 (*b*)), the dissipated energy $E_{dis}$ (FIG. 9 (*c*)), the inactive potential energy $E_d$ due to deposition (FIG. 9 (*d*)), the total converted potential energy $E_{con}$ (i.e., $E_k$+ $E_{dis}$, FIG. 9 (*e*)), and the total unconverted (including both the active and inactive parts) potential energy $E_r$ (i.e., $E_p$+ $E_d$, FIG. 9 (*f*)) over different slopes.

Compared to the absolute values of the kinetic energy (FIG. 8) that differ greatly as the bed slope varies, a consistent overall variation pattern for the proportion of kinetic energy can be identified (FIG. 9 (*a*)). Specifically, three stages can be identified: an initial stage of rapid increase, followed by a relatively stable stage and finally, a decreasing stage. In the initial stage, the kinetic energy proportion $E_k/E_{tot}$ increases to its maximum value (55 % for case 1 at $t$ = 3.2, 51 % for case 2 at $t$ = 3.5, 50 % for case 3 at $t$ = 3.8, 51 % for case 6 at $t$ = 4.3, 54 % for case 7 at $t$ = 6.7). This occurs because the conversion of potential energy to kinetic energy (FIG. 9 (*e*)) at the initial stage (when the collapsing effect dominates) is much more than the energy dissipation (FIG. 9 (*c*)). The different timing for the maximum value of the kinetic energy proportion $E_k/E_{tot}$ is due to the sloping effects. The steeper the bed slope, the stronger the collapsing effect, and thus the later the appearance of the maximum value. The relatively stable behavior of the proportion $E_k/E_{tot}$ after the initial stage indicates a temporary balance between the conversion of potential energy to kinetic energy (FIG. 9 (*e*)) and the dissipation of energies (FIG. 9 (*c*)). This occurs because of the weakening collapsing effect and thus the reduced conversion rate of potential energy to kinetic energy. The final decreasing stage of the proportion $E_k/E_{tot}$ indicates that the energy dissipation overwhelms the conversion of potential energy to kinetic energy. On the one hand, this is due to the further weakening of the conversion, and on the other hand, the increasingly dominant viscous effects that dissipate energies (FIG. 9 (*c*)). Similar to FIG. 9 (*a*) for the kinetic energy proportion $E_k/E_{tot}$, the proportion $E_p/E_{tot}$ of the remaining potential energy also exhibits a consistent variation trend for turbidity currents over different slopes. The potential energy proportion $E_p/E_{tot}$ decreases with time. Two stages can be noted. In the initial collapsing period, the potential energy proportion $E_p/E_{tot}$ decreases very rapidly. Later, it decreases gradually at a considerably reduced rate.

For those turbidity currents moving over a steeper slope, there tends to be relatively less potential energy being converted $E_{con}$ (see FIG. 9 (*e*)), less energy being dissipated $E_{dis}$ (see FIG. 9 (*c*)), clearly more sediment deposition-induced inactive potential energy $E_d$ (see FIG. 9 (*d*)), and slightly more active potential energy $E_p$ (see FIG. 9 (*b*)). These can be explained by the different extent of sediment deposition over different slopes. The results in FIG. 8 demonstrate that a turbidity current over a steeper slope has higher kinetic energy, which implies a higher ability to transport sediment and thus, less sediment deposition. More sediment in suspension leads to high proportion of $E_p$ (FIG. 9 (*b*)). In contrast, the behavior of the inactive potential energy proportion $E_d/E_{tot}(t)$ depends on the quantity and relative location of sediment deposition (see Eq. (10c) for the definition of $E_d$), as well as the initial total energy. On the one hand, sediment deposits mainly near the lock position where the sediment initially stays. This means that the difference of the relative position of the sediment decreases as the dynamic reference plane descends with time. On the other hand, the proportion of the sediment deposition compared to the total amount of sediment is essentially of similar magnitude for depositional turbidity currents. These two aspects lead to a higher proportion of inactive potential energy $E_d/E_{tot}(t)$ over a steeper slope (FIG. 9 (*d*)). The proportion of the remaining potential energy $E_r$ (= $E_p$ + $E_d$) is thus larger (FIG. 9 (*f*)), implying a smaller percentage of the total energy being converted (FIG. 9 (*e*)). Accordingly, the proportion of the dissipated energy that originates from the converted energy is also smaller (FIG. 9 (*c*)).

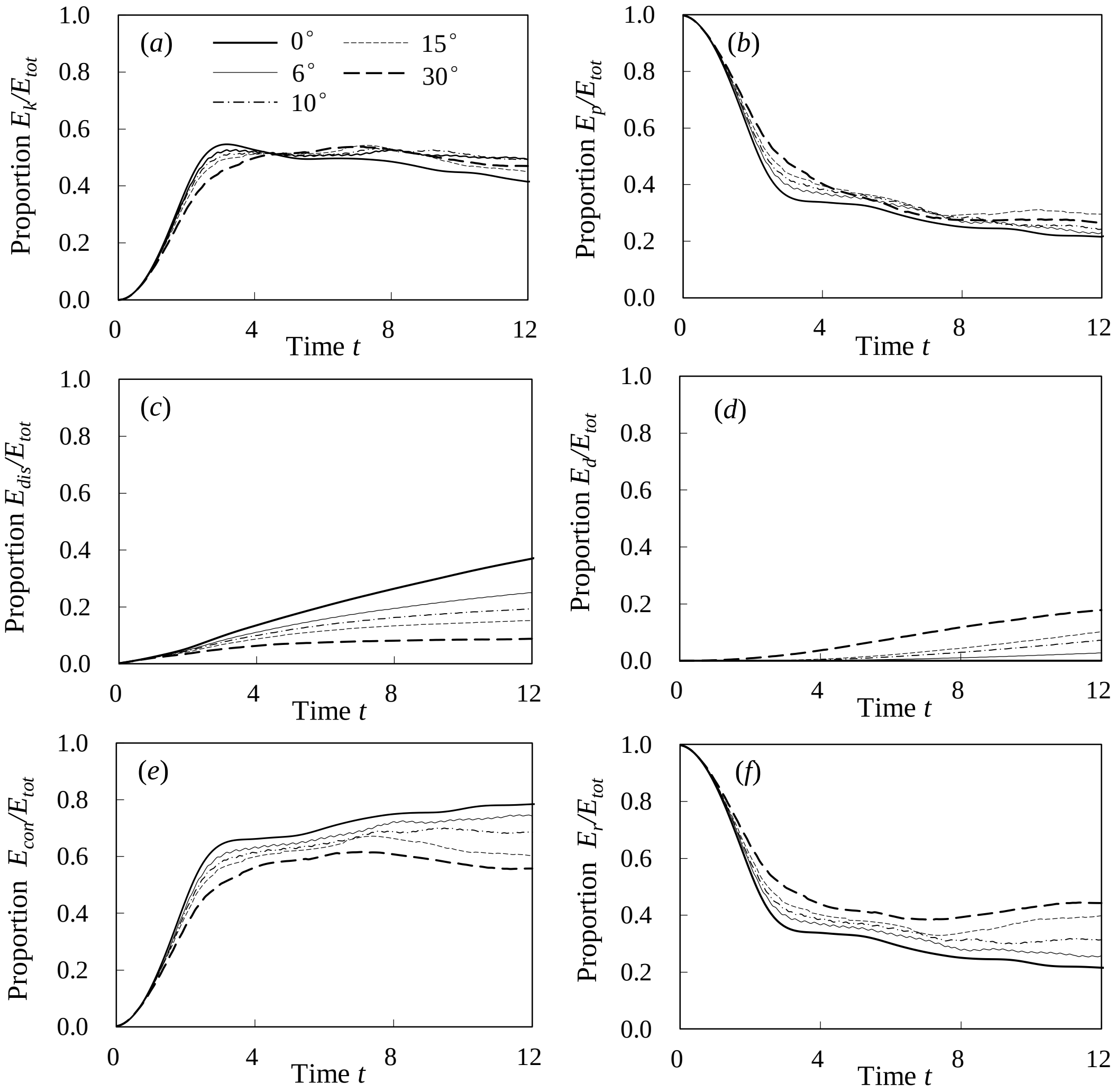

FIG. 9. Time evolutions of the proportion of different energy components: (*a*) $E_k/E_{tot}(t)$, (*b*) $E_p/E_{tot}(t)$, (*c*) $E_{dis}/E_{tot}(t)$, (*d*) $E_d/E_{tot}(t)$, (*e*) $E_{con}/E_{tot}(t)$ and (*f*) $E_r/E_{tot}(t)$ of turbidity currents down different slopes. $Re$ = 3000, $u_s$ = 0.02. These energy terms are normalized by the initial potential energy $E_{tot}(t)$.

### 3.2.2 Influence of the settling velocity on the energy budget

FIG. 10 illustrates the effects of the particle settling velocity on the proportions of these energy components (cases 3, 4, and 5). A turbidity current carrying coarser sediment deposits more sediment, and thus has less along-slope effective gravitation. Accordingly, turbidity currents carrying coarser sediments exhibit a smaller proportion of $E_p$ (FIG. 10 (*b*)) and $E_k$ (FIG. 10 (*a*)) but a larger portion of $E_d$ (FIG. 10 (*d*)). On the other hand, since the presence of the settling velocity contributes to an additional energy loss (Eq. (10d)), a higher percentage of the total energy of the turbidity current with a larger $u_s$ is dissipated (FIG. 10 (*c*)). The significant increase in the dissipated energy also makes the portion of the converted energy larger (FIG. 10 (*e*)). Consequently, a smaller percentage of the total energy remains in the system (FIG. 10 (*f*)).

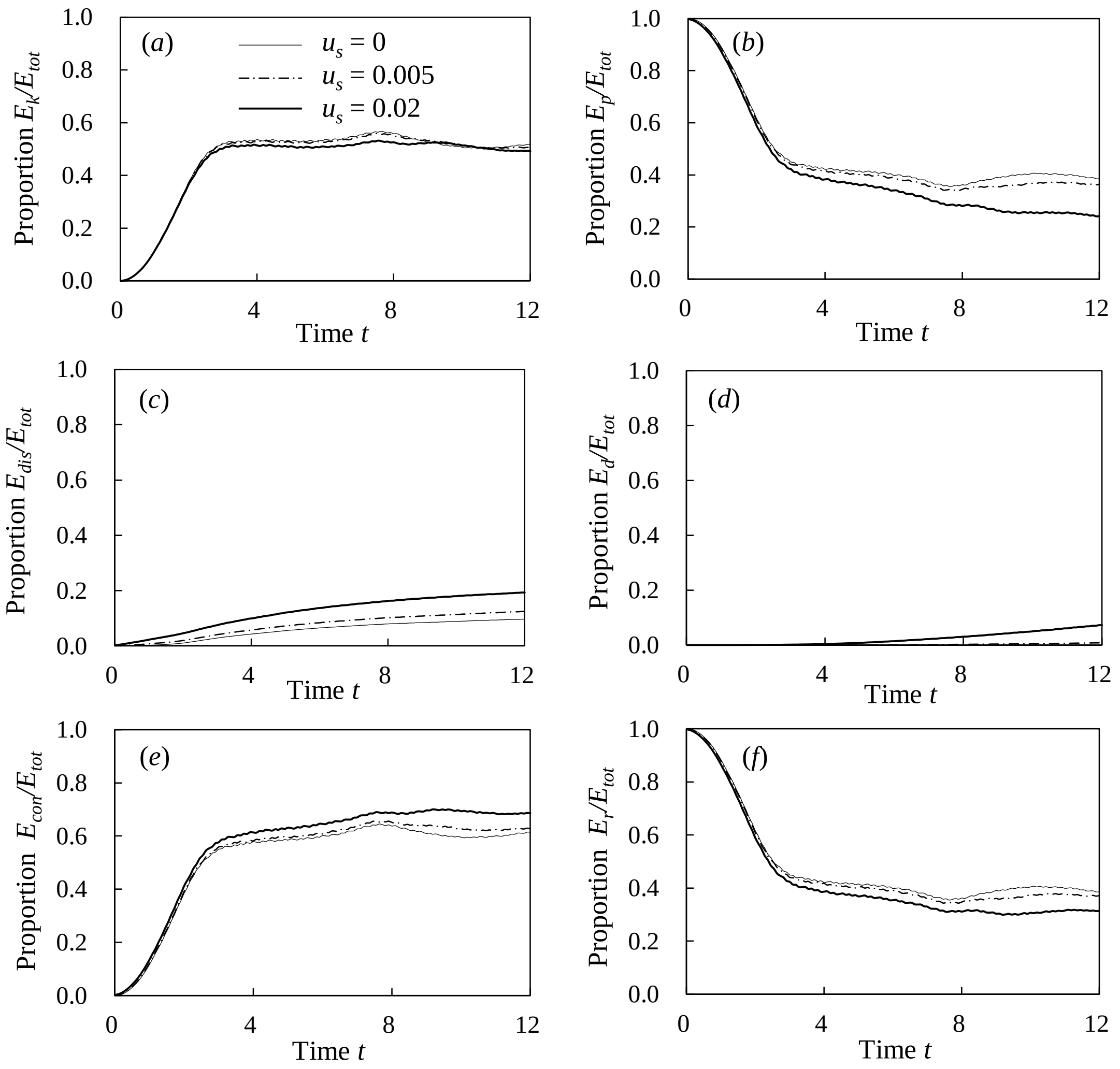

FIG. 10. Time evolutions of the proportion of different energy components: (*a*) $E_k/E_{tot}(t)$, (*b*) $E_p/E_{tot}(t)$, (*c*) $E_{dis}/E_{tot}(t)$, (*d*) $E_d/E_{tot}(t)$, (*e*) $E_{con}/E_{tot}(t)$ and (*f*) $E_r/E_{tot}(t)$ of turbidity currents with different particles down a 10° slope. $Re$ = 3000. These energy terms are normalized by the initial potential energy $E_{tot}(t)$.

### 3.3 Front velocity

The current front velocity is normally used as an index for distinguishing different stages of the turbidity current evolution.

### 3.3.1 Influence of the settling velocity on the front velocity

FIG. 11 (*a*) shows the time evolution of the front velocity of turbidity currents down an intermediate slope of 10° with different settling velocities (cases 3, 4, and 5). From FIG. 11 (*a*), the evolution process of a turbidity current down the intermediate slope can be divided into three distinct stages. It can be seen that, from $t = 0$ to $t \approx 1.4$, the sudden collapse of the dense water leads to a short acceleration stage, during which $U_f$ increases rapidly from zero to the maximum value $U_{fmax} \approx 0.67$. After $t \approx 1.4$, the turbidity current experiences a quasi-constant stage with $U_f$ maintaining a nearly constant value of 0.66. In this stage, a temporary balance is reached between the viscous force and the driving force. Due to entrainment with ambient water, the driving force decreases and thus the current enters the next deceleration stage at approximately $t \approx 8.8$.

When the settling velocity is varied, quantitative differences are seen in FIG. 11 (*b*). The parameter $m_p$ in FIG. 11 (*b*) represents the nondimensional mass of sediments in motion. It can be seen that the higher the settling velocity is, the smaller the current front velocity. This is because a higher settling velocity causes greater sediment settling, which reduces the driving force of the turbidity current and therefore the current front velocity. Nevertheless, the qualitative behaviors, as described above, are consistent. It is because only a small part of the total sediment is deposited, as shown in FIG. 11 (*b*). Specifically, all the sediments are suspended for the case of $u_s = 0$. When $u_s$ increases to 0.02, over 80 % of the total sediments are still suspended at $t = 12$.

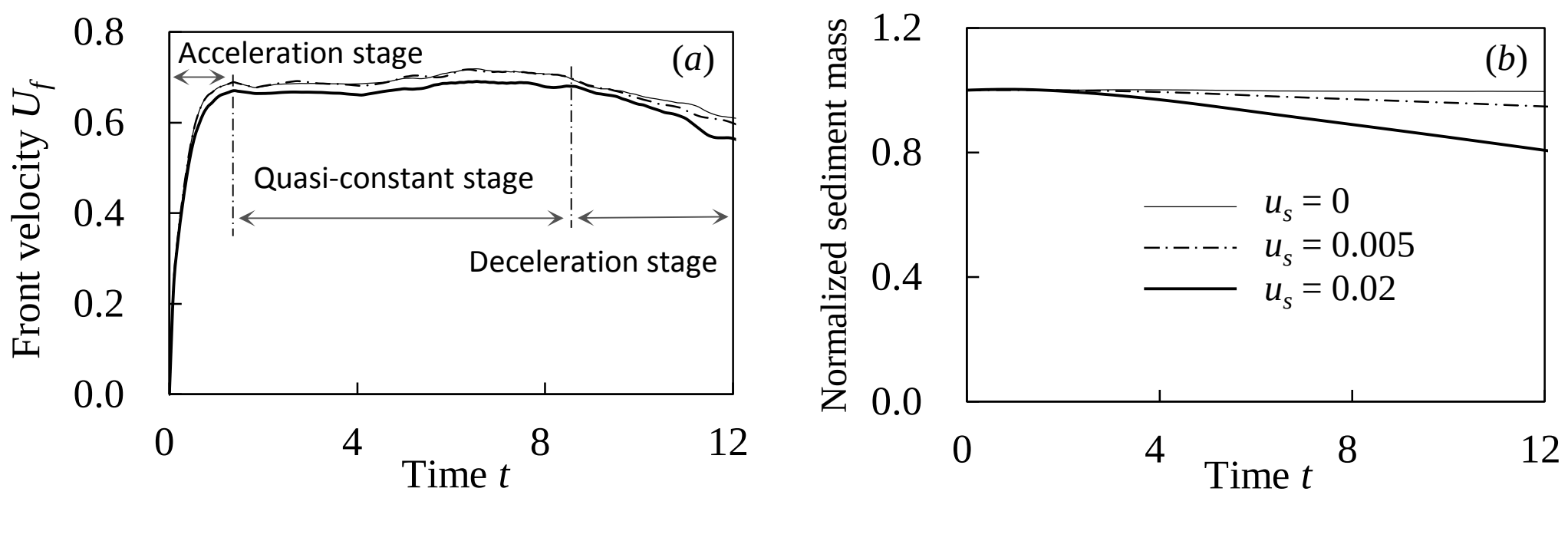

FIG. 11. Time evolutions of (*a*) the front velocity of turbidity currents with different settling velocities down a 10° slope and (*b*) the mass of the suspended particles (i.e., the proportion of sediment that remains suspended). *Re* = 3000.

### 3.3.2 Influence of the slope angle on the front velocity

FIG. 12 shows time evolutions of the front velocity of turbidity currents down different slopes (cases 1, 2, 3, 6, and 7). In all the cases, the current front velocity $U_f$ experiences a rapid increase stage before $t \approx 1.4$. During this stage, $U_f$ is not greatly influenced by $\theta$ due to the dominance of the collapsing effect. Afterwards, turbidity currents tend to behave differently depending on the magnitude of the bed slope. The quasi-constant stage is absent for a turbidity current over a horizontal bottom. The current transits from the accelerating stage to the decelerating stage directly. Specifically, its $U_f$ decreases by approximately 20 % from $t \approx 1.4$ to $t \approx 9$, and approximately 18 % from $t \approx 9$ to $t = 12$. For those over an intermediate slope (e.g., 6° or 10°), the quasi-constant stage is retained, with $U_f \approx 0.65$ from $t \approx 1.4$ to $t \approx 8.6$ for $\theta = 6°$ and $U_f \approx 0.68$ from $t \approx 1.4$ to $t = 8.4$ for $\theta = 10°$. After $t \approx 8.6$, the front velocity starts to decrease. For those over a steep slope, the quasi-constant stage is also missing. Instead, the current has an extended accelerating stage, though its accelerating rate is much reduced compared to the initial rapid accelerating stage. For the cases of $\theta = 15°$, the accelerating rate of the front velocity is approximately 0.012 during the extended accelerating stage (from $t \approx 1.4$ to $t \approx 7.1$), compared to 0.503 during the initial rapid accelerating stage ($t < 1.4$). For the case of $\theta = 30°$, this trend is more obvious.

At the initial period, the sudden removal of the lock leads to a rapid collapse of dense fluid. At the final period, the density contrast is small, and the turbidity current is mainly dominated by the viscous effect. Consequently, turbidity currents down different slopes all have the first rapid acceleration stage and the final deceleration stage. The along-slope effective gravitation that serves as the positive effect increases with $\theta$. The interplay of the positive (effective along-slope gravitation) and negative effect (viscous force) results in the presented behavior of the front velocity in the second stage. For a relatively small slope (e.g., $\theta = 0$), $U_f$ slightly decreases because the negative effect exceeds the positive effect. With the increase in $\theta$, a temporary balance is reached between the positive and negative effects. The second phase thus turns into a quasi-constant stage on an intermediate slope (e.g., $6° \leq \theta \leq 10°$), during which $U_f$ is characterized by a nearly constant value. When $\theta$ becomes larger, the positive effect gradually surpasses the negative effect. The quasi-constant stage is replaced by an accelerating process (e.g., $\theta = 30°$), during which the acceleration of the turbidity current increases with $\theta$.

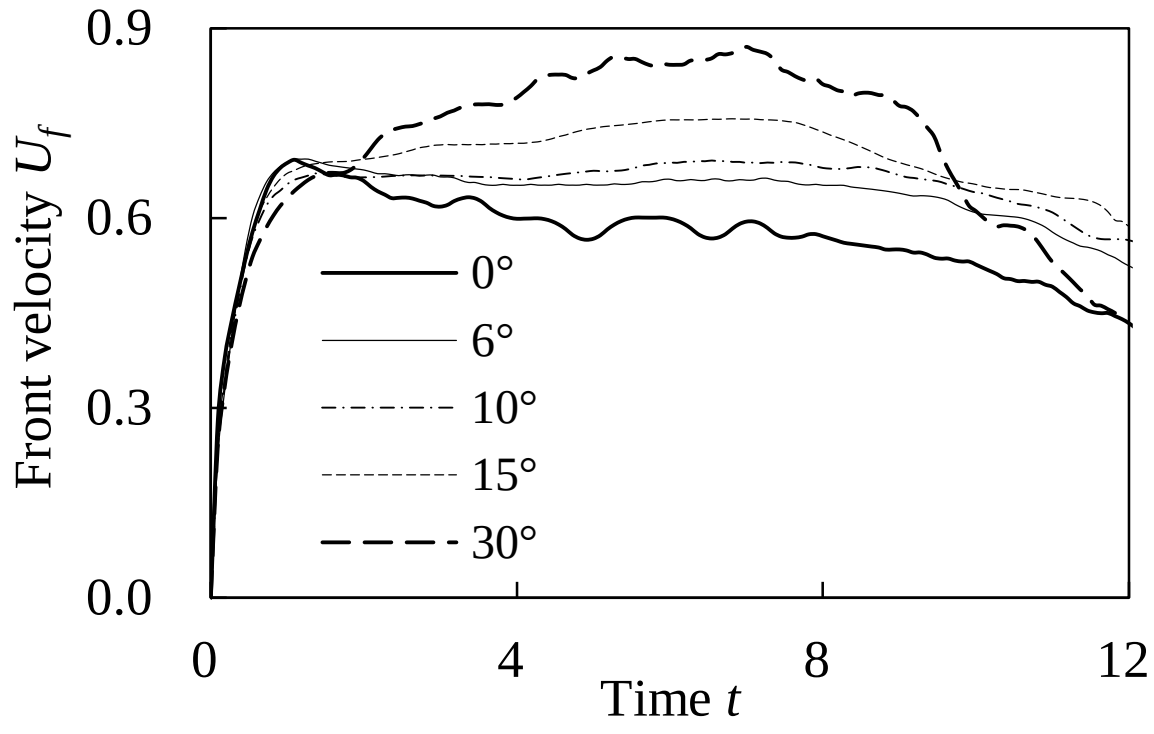

FIG. 12. Time evolutions of the front velocity of turbidity currents down different slopes. $Re$ = 3000.

### 3.4 Entrainment ratio

When a turbidity current propagates downstream, it entrains ambient water and dilutes the turbid water. This phenomenon is normally quantified using water entrainment ratios, which can be defined as follows [52]

$$E_{bulk,n} = \frac{W_{bulk,n}}{U_{f,n}}, \tag{11a}$$

$$E_{inst,n} = \frac{W_{inst,n}}{0.5(U_{f,n} + U_{f,n-1})}, \tag{11b}$$

where $E_{bulk,n}$ and $E_{inst,n}$ represent the bulk and instantaneous water entrainment ratios, respectively; the subscript *bulk* indicates a time-averaged bulk value; the subscript *inst* indicates a transient value; the subscripts $n$ and $n$ – 1 are time step indicators corresponding to $t_n$ and $t_{n-1}$; $U_{f,n}$ and $U_{f,n-1}$ are the front velocities at $t = t_n$ and $t = t_{n-1}$; and $W_{bulk,n}$ and $W_{inst,n}$ represent the bulk and instantaneous water entrainment velocity, which are defined as

$$W_{bulk,n} = \frac{Q_{bulk,n}}{S_n}, \tag{12a}$$

$$W_{inst,n} = \frac{Q_{inst,n}}{0.5(S_n + S_{n-1})}, \tag{12b}$$

where $Q_{bulk,n}$ and $Q_{inst,n}$ are the effective bulk and instantaneous water entrainment discharge and $S_n$ and $S_{n-1}$ are the area of the upper interface at $t = t_n$ and $t = t_{n-1}$. A first-order relation proposed by Ottolenghi *et al.* [17,53] is used to compute the area $S_n = (h + l + X_{f,n}) \times 1$, where $X_{f,n}$ is the front position at $t = t_n$. The parameters $Q_{bulk,}$ and $Q_{inst,n}$ can be estimated using the principle of mass conservation, i.e., an increase in turbidity current volume is due to water entrainment $\Delta V = Q\Delta t$ . Therefore,

$$Q_{bulk,n} = \frac{V_n - V_0}{t_n}, \qquad (13a)$$

$$Q_{inst,n} = \frac{V_n - V_{n-1}}{t_n - t_{n-1}}, \qquad (13b)$$

where $V_0$, $V_n$ and $V_{n\text{-}1}$ are the volume of the turbidity current at $t$ = 0, $t = t_n$ and $t = t_{n\text{-}1}$. The current volume $V$ at any given time is computed as 1 × $A$, where $A$ is the area with the particle concentration $c$ > 0.01, i.e.,

$$V = 1\times A = 1\times\int_0^H\int_0^L \phi_{(c)}dxdz, \quad \text{with}\begin{cases}\phi_{(c)} = 1, & 0.01 < c\\ \phi_{(c)} = 0, & c \le 0.01\end{cases}. \qquad (14)$$

The choice of the criterion $c$ > 0.01 is based on a sensitivity analysis that shows the current volume changes only slightly when the concentration $c$ varies between 0.01 and 0.05.

#### 3.4.1 Influence of the slope angle on water entrainment

FIG. 13 shows time evolutions of the entrainment ratio $E_{bulk}$ and $E_{inst}$ of turbidity currents down different slopes (cases 1, 2, 3, 6, and 7). It appears that the bed slope does not have an obvious effect on the overall tendency of both the bulk and instantaneous entrainment ratios. Turbidity currents over a steeper bed tend to have a greater front velocity (FIG. 12) and at the same time, a larger mixed region (FIG. 7), or alternatively, a larger volume of entrained water. The increase in both the numerator (water entrainment velocity) and the denominator (front velocity) leads to a subtle effect on the entrainment ratios. The following is devoted to the overall trend of the entrainment ratios. Both $E_{bulk}$ and $E_{inst}$ assume the greatest values at the very beginning, which is due to the engulfing of water accompanying the initial collapse and the formation of the head [54]. Later, both $E_{bulk}$ and $E_{inst}$ decrease rapidly with time, as the collapsing effects diminish during the acceleration stage (from $t$ = 0 to $t$ ≈ 1.4). Afterwards, water entrainment is due to the Kelvin-Helmholtz instabilities and turbulent billows developed at the upper interface (FIG. 5) [17,54]. During this stage, both $E_{bulk}$ and $E_{inst}$ take on a relatively small value of approximately 0.05. Compared to the relatively stable values of the bulk entrainment ratios (FIG. 13 ($a$)), the instantaneous entrainment ratio fluctuates at approximately 0.05. The fluctuation of $E_{inst}$ is essentially related to the emergence and disappearance of the turbulent billows

(see FIG. 5) [17].

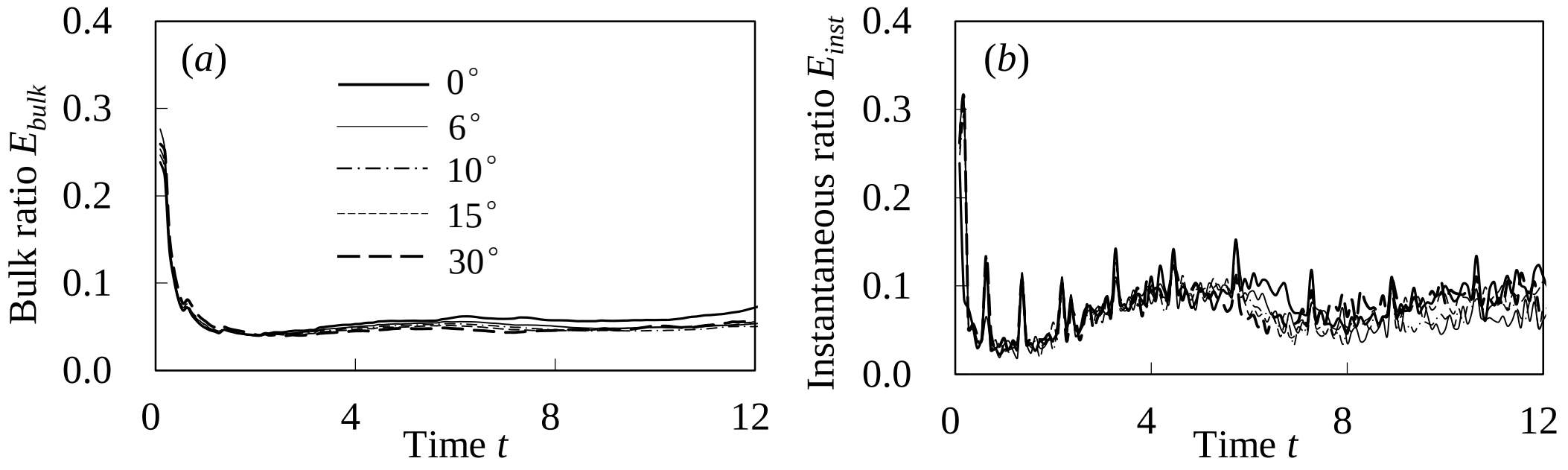

FIG. 13. Time evolutions of (*a*) bulk entrainment ratios, and (*b*) instantaneous entrainment ratios of turbidity currents down different slopes. *Re* = 3000. $u_s$ = 0.02.

### 3.4.2 Influence of the settling velocity on water entrainment

Here, the effects of the settling velocity on the entrainment ratio are analyzed. The presence of sediment leads to two important effects on the water entrainment. Sediment deposition may lead to a weaker density stratification at the upper interface on the one hand [31], and a reduced driving force and thus a smaller front velocity on the other hand. The entrainment effect is enhanced by a weaker interface stratification [31] and is decreased by a smaller front velocity. FIG. 14 (*a*) and (*b*) show time evolutions of $E_{bulk}$ and $E_{inst}$ of turbidity currents with different $u_s$ down a 10° slope. Also shown in FIG. 14 (*c*) and (*d*) are time evolutions of $E_{bulk}$ and $E_{inst}$ of turbidity currents down a flat bed. The entrainment ratios of turbidity currents with different $u_s$ have almost the same development tendency and value, which means that the two opposite effects on the entrainment ratio offset each other. As the settling velocity has little effect on $E_{bulk}$ and $E_{inst}$ of a current down a slope, the value of the entrainment ratio measured from a compositional gravity current may be applied in the case of a turbidity current, at least in the slope within the range between 0° and 10°.

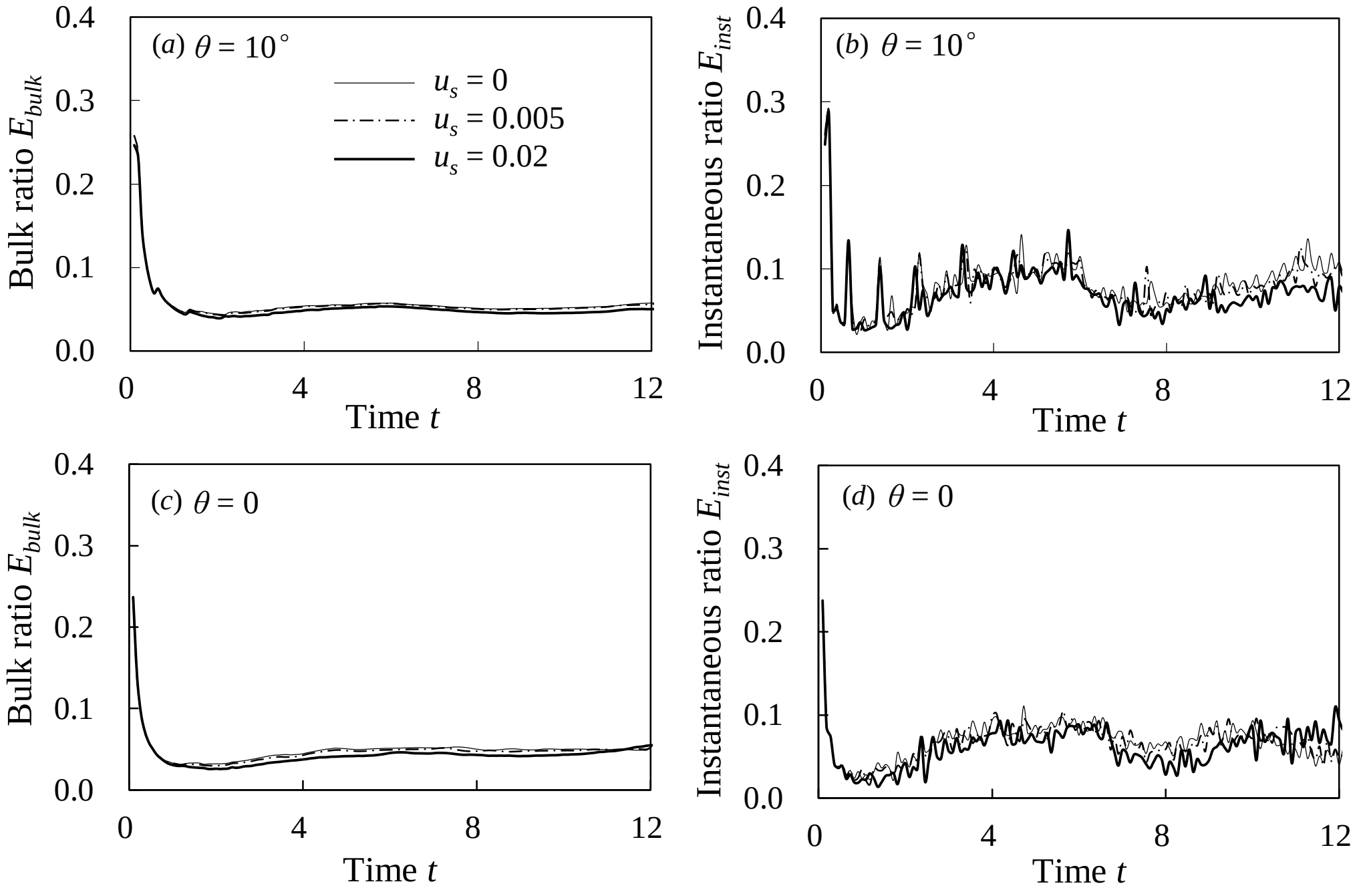

FIG. 14. Time evolutions of (*a*) bulk entrainment ratios and (*b*) instantaneous entrainment ratios of turbidity currents with different particles down a 10° slope and (*c*) bulk entrainment ratios and (*d*) instantaneous entrainment ratios of turbidity currents on a flat bed. *Re* = 3000.

### 3.5 Comparisons with 3D results

Two additional 3D simulations are run to make comparisons with the present 2D simulations. The width in the spanwise direction in 3D cases is set at 2, with a non-slip and non-flux boundary condition applied at two lateral walls. The other simulation setups are the same as the 2D cases (on a flat bed, see case 1 in Table 2; down a 10° slope, see case 2 in Table 2). The comparisons are conducted towards the front location and energy term, the mixed region, and the bulk entrainment ratio, as shown in FIG. 15. It is evident that the predictions from the two approaches exhibit some quantitative differences due to ignoring the wall effects in the 2D cases. However, these quantitative differences are very much limited. Both 2D and 3D DNS approaches provide qualitatively similar trends for these parameters. Since this study aims to reveal the effects of the particle settling velocity and bed slope on the bulk properties of particle-laden gravity currents (e.g., the longitudinal variation in front velocity, energy budget and entrainment effect), instead of resolving the 3D structures, the 2D DNS approach is considered appropriate for the purpose of the present study. Although using 2D DNS would veritably sacrifice some extent of the accuracy (e.g., the 3D structures developed at the later stage of the current), it does not affect the understanding obtained by this work.

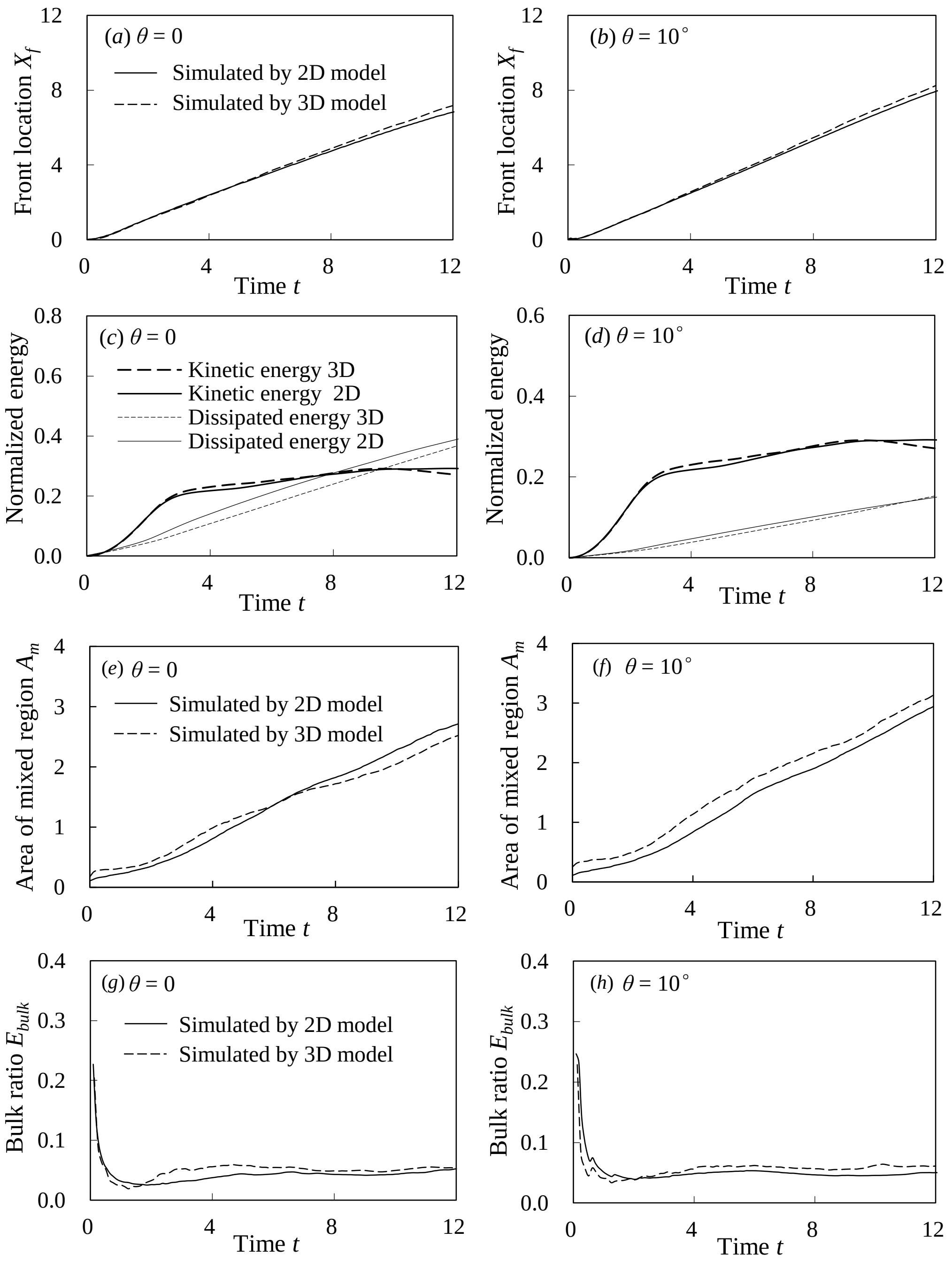

FIG. 15. Comparisons between 2D DNS and 3D DNS in terms of front location, energy term, mixed region, and bulk entrainment ratio.

# 4 Conclusions

In this paper, two-dimensional direct numerical simulations are implemented for lock-exchange turbidity currents down different slopes ($\theta \leq 30°$). The obtained high-resolution numerical results are used to analyze the effects of the bed slope and the particle settling velocity on the dynamics in terms of the energy budget, front velocity, and water entrainment ratio in 12 nondimensional time units.

Initially, a reservoir of turbid water is placed on the left end of the computational domain. The potential energy of this reservoir of turbid water provides the energy sources of the later-formed turbidity current. Once the current starts to move in response to lock-release, potential energy is converted into kinetic energy ($E_k$). At the same time, some part of the kinetic energy is dissipated; some part of the potential energy also becomes unconvertible or inactive ($E_d$) as suspended particles deposit on the topography. Our simulation shows that during the initial period, the collapse of the dense fluid introduces a very fast energy conversion process of potential energy into kinetic energy $E_k$, with only a small part of the kinetic energy dissipated. Later, a leading head of the turbidity current is formed, with Kelvin–Helmholtz instabilities and turbulent billows generated at the upper interface, which promote an entrainment effect with the ambient water. After a first fast energy conversion process, the energy conversion rate slows down. The behavior of $U_f$ after the acceleration stage is determined by the interplay of the driving force and the viscous force.

The main highlights and conclusions of this study are as follows:

1. A dynamic reference plane, which is defined at the lowest point of the turbidity current and descends with time, is proposed to analyze the energy budget of a turbidity current down a slope. By this definition, the total initial energy of the turbidity current that increases with time can serve as a good energy scale to analyze the temporal variation in the proportions of different energy components. In contrast to the static reference plane defined previously (at the position of the initial turbid water), the dynamic reference plane can avoid the negative values of the energy terms.

2. The absolute values of the energy components are diverse when the slope angle and the settling velocity are different. These diversities can be greatly reduced by normalizing the absolute values with the total initial energy. The normalized kinetic energy tends to first quickly increase, then remain nearly constant, and finally decrease. The normalized potential energy continuously decreases, with an initially great decelerating rate.

3. The evolution process of a turbidity current down a slope can be divided into a first acceleration stage, a second stage during which its behavior is determined by the slope angle, and a third deceleration stage. In the second stage, $U_f$ changes from a slight decrease on a relatively slight slope (e.g., $\theta$ = 0), to a quasi-constant state on an intermediate slope (e.g., $\theta$ = 10°), and then an increasing process on a relatively steep bed (e.g., $\theta$ = 30°). The turbidity current with a larger $u_s$ has a smaller front velocity.

4. Both bed slope and sediment settling do not show an appreciable effect on the water entrainment ratios. During the first acceleration stage, both the bulk and instantaneous entrainment ratios ($E_{bulk}$ and $E_{inst}$) acquire the greatest values due to the engulfing of ambient water in the collapsing process of the particle-laden fluid. Later, $E_{inst}$ fluctuates with time intensely while $E_{bulk}$ maintains a nearly constant value of approximately 0.05.

## Acknowledgements

This work was partially supported by the National Natural Science Foundation of China (11672267; 11772300), Natural Science Foundation of Zhejiang Province (LR16E090001), Fundamental Research Funds for the Central Universities (2017XZZX001-02A, 2018QNA4049) and Research Funding of Shenzhen City (JCYJ20160425164642646).

## Appendix:

The following symbols are used in this paper. For the symbols with both dimensional and non-dimensional forms, only the dimensional form is listed (the corresponding non-dimensional form is without a $\hat{}$ sign).

$\hat{H}$ : the flume height;

$\hat{L}$ : the flume length;

$\hat{l}$ : the length of the reservoir storing the initial turbid water;

$\hat{h}$ : the height of the reservoir storing the initial turbid water;

$\hat{c}_0$ : the initial sediment concentration;

$\theta$: the slope angle;

$\hat{x}$ : the horizontal direction;

$\hat{z}$ : the vertical direction;

$\hat{x}_f$ : the front location of the turbidity current;

$\hat{u}_b$ : the buoyancy velocity;

$Re$: Reynolds number;

$\hat{\nu}$ : the kinematic viscosity of the fluid;

$\hat{\rho}_p$ : the density of the sediment;

$\rho$: the density of fluid;

$g$ : gravitational acceleration;

$u$: the fluid velocity;

$\hat{U}_f$ : the front velocity of the turbidity current;

$\hat{t}$ : the time;

$\hat{p}$ : the pressure;

$\hat{u}_s$ : the settling velocity of the sediment;

$\hat{c}$ : the sediment concentration;

$Sc$; the Schmidt number;

$\hat{k}$ : the diffusivity of the sediment concentration field;

$\hat{d}_p$ : the sediment diameter;

$z_\perp$ : the direction of the normal vector on the bottom surface;

$f$: the virtual force;

$\Delta t$: the time step;

$\eta$ : the percent of the volume of the solid phase;

$\Delta x$: the spatial step;

$A_m$: the mixed region of the turbidity current;

$\phi_{(c)}$ : a defined parameter changing with sediment concentration;

$E_{p0}$ : the initial potential energy of the turbid water;

$E_{con}$ : energy which has been converted to kinetic energy;

$E_d$ : the inactive potential energy;

$E_p$ : the remaining active potential energy;

$E_k$: the kinetic energy;

$E_{dis}$ : the dissipated energy;

$E_r$: potential energy that is not converted;

$E_d$: inactive potential energy due to sediment deposition;

$z_d$ : the reference plane;

$z_b$: the bed elevation;

$m_d$: the deposited sediments;

$\varepsilon_f$ and $\varepsilon_s$: energy dissipation rate;

*s*: the rate-of-strain tensor;

$E_{tot}$: the total energy;

$E_{bulk}$: the bulk water entrainment ratio;

$E_{inst}$: the instantaneous water entrainment ratio;

$W_{bulk}$: the bulk and instantaneous water entrainment velocity;

$W_{inst}$: the instantaneous water entrainment velocity;

$Q_{bulk}$: the effective bulk water entrainment discharge;

$Q_{inst}$: the effective instantaneous water entrainment discharge;

*S*: the area of the upper interface;

*V*: the volume of the turbidity current;

*A*: the area where the sediment concentration is larger than 0.01.